\documentclass[aps,epsfig,manuscript,showpacs]{revtex4}

\usepackage{amsmath,amssymb,graphicx}
\usepackage{epsfig}
\usepackage{graphicx}
\usepackage{enumerate}
\newcommand{\beq}{\begin{equation}}
\newcommand{\eeq}{\end{equation}}

\newcommand{\bea}{\begin{eqnarray}}
\newcommand{\eea}{\end{eqnarray}}

\begin{document}

\title{Heat flow in nonlinear molecular junctions}

\author{Dvira Segal}
\affiliation{Department of Chemical Physics, Weizmann
Institute of Science, 76100 Rehovot, Israel}

\begin{abstract}

We investigate the heat conduction properties of molecular
junctions comprising anharmonic interactions.
We find that nonlinear interactions can lead to novel phenomena:
{\it negative differential thermal conductance} and heat rectification.
Based on analytically solvable models we derive an expression for
the heat current that  clearly reflects the interplay between
anharmonic interactions, strengths of coupling to the thermal
reservoirs, and junction asymmetry.
This expression indicates that negative differential thermal
conductance shows up when the molecule is strongly coupled to the
thermal baths, even in the absence of internal molecular
nonlinearities. In contrast, diode like behavior is expected for a
highly anharmonic molecule with an inherent structural asymmetry.
Anharmonic interactions are also necessary for manifesting
Fourier type transport. We briefly present an extension of our
model system that can lead to this behavior.

\end{abstract}

 \maketitle

\section{Introduction}

Understanding and controlling heat flow in nanoscale structures
is of interest both from the fundamental aspect \cite{Schwab1} and for device
applications \cite{Cahill,Shi,Nanos,Inst}. The influential role of quantum
effects and geometrical constrictions in low dimensional systems
often results in fundamentally interesting behavior \cite{Blencowe}.
Recent theoretical and experimental studies demonstrated that the
thermal transport properties of nanowires can be very different
from the corresponding bulk properties \cite{Rego,Schwab}. In the
low temperature {\it ballistic} regime the phonon thermal
conductance of a one-dimensional (1D) quantum wire is quantized,
with $g=\pi^2k_B^2T/3h$  as the universal quantum conductance unit
\cite{Rego}, where $k_B$ and $h$ are the Boltzmann and Planck
constants, respectively, and $T$ is the temperature. Reflections
from the boundaries and disorder in the wire can be further
treated by considering a Landauer type expression for the heat
current ($\hbar=1$) \cite{Rego,Ciraci,Segalheat}
\beq J=\int d\omega \omega\mathcal{T} (\omega)\left[
n_L(\omega)-n_R(\omega)\right].
\label{eq:Landauer}
\eeq
This relationship describes energy transfer between two (left
($L$) , right ($R$)) thermal reservoirs maintained at equilibrium
with the temperatures $T_L$ and $T_R$, respectively, in terms of
the {\it temperature independent} transmission coefficient
$\mathcal T(\omega)$ for phonons of frequency $\omega$. Here
$n_K(\omega)=\left(e^{\beta_K\omega}-1 \right)^{-1}$;
$\beta_K=1/k_BT_K$, $(K=L,R)$, is the Bose-Einstein distribution
characterizing the reservoirs. This expression assumes the absence
of inelastic scattering processes, and the two opposite phonon
flows of different temperatures are out of equilibrium with each
other. This leads to an {\it anomalous} transport of heat, where
(classically) the energy flux is proportional to the temperature
difference, $\Delta T=T_L-T_R$, rather than to the temperature
gradient, $\nabla T$, as asserted by the Fourier law of
conductivity,
\beq
J= - \mathcal{K}\mathcal{A}\nabla T.
\label{eq:Fourier}
\eeq
In this equation $\mathcal A$ is the cross section area normal to
the direction of heat propagation and $\mathcal K$ is the
coefficient of thermal conductivity. An outstanding problem in
statistical physics is to find out the necessary and sufficient
conditions for attaining this {\it normal} (Fourier) law of  heat
conductivity in low dimensional systems
\cite{Livi,Joel,Prosen,LiWang,Gruber,Schr}. Among the crucial
requirements explored is that the molecular potential energy
should constitute strong anharmonic interactions.

Heat conductance experiments on short molecules or highly ordered
structures provide results consistent with the Landauer
expression. A micron length individual carbon nanotube conducts
heat ballistically without showing signatures of phonon-phonon
scattering for temperatures up to 300 K
\cite{MajumdarCNT,ChiuCNT}. Intramolecular vibrational energy flow
in bridged azulene-anthracene compounds could be
explained by assuming ballistic energy transport in the chain
connecting both chromophores \cite{Schwarzer}. In contrast,
calculations of heat flow through proteins show substantial
contribution of anharmonic interactions leading to an enhancement
of the energy current in comparison to the (artificial) purely
harmonic situation \cite{Leitner}.

Anharmonic (nonlinear) interactions are  also a tool  for  {\it
controlling} heat flow in molecular junctions with potential
technological applications, e.g. a thermal diode
\cite{Casati1,Casati2,Rectif} and a thermal transistor
\cite{Casatixxx}. We have recently demonstrated that when
nonlinear interactions govern heat conduction, the heat current is
asymmetric for forward and reversed temperature biases, provided
the junction has some structural asymmetry \cite{Rectif}.

In this paper we generalize the model developed in Ref.
\cite{Rectif}, and present a comprehensive analysis of the heat
conduction properties of molecular junctions taking into account
anharmonic interactions in the system. We discuss the influence of
the following effects on  heat flow through the junction: (i)
interparticle potential, specifically the degree of molecular
anharmonicity, (ii) molecule-thermal reservoirs contact
interactions, and (iii) junction asymmetry with respect to the $L$
and $R$ ends. We derive an exact analytic expression for the heat
current that clearly reflects the role of each of these factors in
determining phonon dynamics.
%
More specifically, we analyze the  necessary conditions for
demonstrating negative differential {\it thermal} conductance
(NDTC) and diode like behavior. We also follow the transition from
the elastic Landauer formula to the Fourier law of conduction as
anharmonic interactions are turned on.

The paper is organized as follows: Section II presents our model
system. Section III begins with a fully harmonic model and shows
that it satisfies the Landauer formula. We then proceed and show
that an asymmetric anharmonic molecule linearly coupled to thermal
reservoirs,  can rectify heat. Section IV further presents
 strong coupling models that exhibit
 NDTC. Under additional conditions, the nonlinear
models extended to a $l$ sites system satisfy the Fourier law of
conductivity as explained in Section V. Section VI provides
concluding remarks.

\section{Model}
The model system consists a molecular unit connecting two thermal
reservoirs left ($L$)  and right ($R$) of inverse temperatures $\beta_L=T_L^{-1}$
and $\beta_R=T_R^{-1}$ respectively. Henceforth we take the
Boltzmann constant as $k_B=1$.
The general Hamiltonian includes three contributions: the
molecular part ($M$), the two reservoirs ($B$) and the system-
bath interaction ($MB$)
\beq
H= H_M +H_B +H_{MB}.
\label{eq:HT}
\eeq
%
For simplicity we assume that heat
transfer is dominated by a specific single mode.
The molecular term in the Hamiltonian is therefore given by
\beq
H_M=\sum_{n=0}^{N-1}E_n |n\rangle \langle n|; \,\,
E_n=n\omega_0,
\label{eq:H1M} \eeq
where $\omega_0$ is the frequency of the molecular oscillator
($\hbar\equiv 1$). We shall consider two situations: harmonic
model, and a two-level system (TLS) that simulates a highly
anharmonic vibrational mode. For a harmonic molecule $N$ is taken
up to infinity. Strong anharmonicity is enforced by limiting $n$
to $0,1$. The molecular mode is coupled either linearly (weakly)
or nonlinearly (strongly) to the $L$ and $R$ thermal baths
represented by sets of independent harmonic oscillators
\beq
H_B=H_L+H_R; \   \ H_K=\sum_{j\in K}{\omega_j a_j^\dagger
a_j} ;\ \ K=L,R.
\label{eq:H1B}
\eeq
$a_j^{\dagger}$, $a_j$ are boson creation and annihilation
operators associated with the phonon modes of the harmonic baths.
The $L$ and $R$ thermal baths are not coupled directly, only
through their interaction with the molecular mode.
We use the following model for the molecule-reservoirs interaction
\bea H_{MB}&=&\sum_{n=1}^{N-1} \left(B|n-1\rangle \langle n|
+B^{\dagger}|n\rangle \langle n-1| \right)\sqrt{n},
\label{eq:H1MB}
\eea
where $B$ are bath operators. This model assumes that transitions
between molecular levels occur due to the environment excitations.
Note that in general this interaction does not need to be additive
in the thermal baths, i.e. we may  consider situations in which $B
\neq B_L+B_R$, see Section IV.

The probabilities $P_n$ to occupy the $n$ state of the molecular
oscillator satisfy the master equation
\bea
 \dot P_n&=&-\left[ n k_d + (n+1) k_u  \right] P_n
\nonumber\\
&+&(n+1)k_dP_{n+1}+n  k_uP_{n-1},
\label{eq:master}
\eea
where the occupations are normalized $\sum_n P_n=1$. Within second
order perturbation theory the rates are given by
\bea
 k_d&=& \int_{-\infty}^{\infty} d\tau e^{i \omega_0 \tau}
\langle B^{\dagger}(\tau) B(0) \rangle,  \,\,\,\
\nonumber\\
k_u&=& \int_{-\infty}^{\infty} d \tau e^{-i \omega_0 \tau} \langle
B(\tau) B^{\dagger}(0) \rangle,
\label{eq:kdku}
\eea
where the average is done over the baths thermal distributions,
irrespective of the fact that it may involve two distributions of
different temperatures. In Section IV we demonstrate that these
rates also apply in the strong molecule-baths interaction limit.

A useful concept in the following discussion is the notion of an effective
molecular temperature. It can be defined through the relative
population of neighboring molecular levels
\beq T_{M}\equiv -\frac{\omega_0}{\log (P_{n+1}/P_n)}.
\label{eq:TM}
\eeq
At steady state this ratio does not depend on $n$, see
Eq.~(\ref{eq:master}). We show below that the molecular temperature
is given in terms of the reservoirs temperatures weighted by the
molecule-baths coupling strengths.

Given the reservoirs temperatures $T_L$ and $T_R$, we can define
two other related parameters: the temperature difference $\Delta
T=T_L-T_R$ and the average temperature $T_a=(T_L+T_R)/2$. The
temperature difference can be experimentally imposed in various ways. Here we
consider two situations: We may fix the temperature at the left
reservoir while varying the temperature at the right side,
\bea
{\rm (A)} \,\,\, T_L&=&T_s
\nonumber\\
T_R&=&T_s-\Delta T. \label{eq:modelA} \eea
For the same temperature difference we can also build a
symmetric situation where the temperatures of both reservoirs are
equally shifted,
\bea
{\rm (B)} \,\,\, T_L&=&T_s+\Delta T/2
\nonumber\\
T_R&=&T_s-\Delta T/2.
\label{eq:modelB}
\eea
The main difference between these two situations is that the
average temperature is decreasing steadily with $\Delta T$
in the first case, while it is  constant ($T_s$) in (B).
We will show below that these boundary conditions determine
the effective {\it molecular}
temperature which implies on the conduction properties of the system.

Next we present the model Hamiltonians in the weak and strong
molecule-bath interaction limits for either purely harmonic or a
TLS molecular mode, and discuss the implications on the junction
thermal conductance.


\section{Weak system-bath coupling}
We begin by analyzing the heat conduction properties of a molecule
coupled {\it linearly} to two thermal reservoirs of different
temperatures \cite{Rectif}. The Hamiltonian is given by Eqs.
(\ref{eq:HT})-(\ref{eq:H1MB}) with linear (harmonic)
system-bath interactions
\bea H_{MB}&=&\sum_{n=1}^{N-1} \left(B|n-1\rangle \langle n|
+B^{\dagger}|n\rangle \langle n-1| \right)\sqrt{n};
\nonumber\\
B&=&B_L+B_R, \label{eq:HMB} \eea
where the bath operators  $B_K$ satisfy
\bea
B_K&=& \sum_{ j \in K}{\bar{\alpha}_j x_j}; \ \
\nonumber\\
x_j&=&(2\omega_j)^{-1/2}({a}_j^{\dagger}+a_j) ;\ \ K=L,R.
\label{eq:H1in}
\eea
In the present linear coupling model, no correlations persist
between the thermal baths and the rate constants (\ref{eq:kdku})
are additive in the $L$ and $R$ baths
\bea
k_d&=&k_L+k_R
\nonumber\\
k_u&=&k_L e^{-\beta_L \omega_0}+k_R
e^{-\beta_R \omega_0},
\label{eq:rate1}
\eea
with
\beq k_K=\Gamma_K(\omega_0) (1+n_K(\omega_0)) ; \ \ K=L,R.
\label{eq:rate2} \eeq
Here $n_K(\omega)=\left( e^{\beta_K\omega}-1\right)^{-1}$,
$\Gamma_K(\omega)= \frac{\pi}{2m\omega^2} \sum_{j \in K}
\alpha_j^2\delta(\omega-\omega_j)$ and $\alpha_j=\bar\alpha_j
\sqrt{2 m\omega_0}$ \cite{Rectif}, where $m$ and $\omega_0$ are
the molecular oscillator mass and frequency respectively.

The heat conduction properties of this model are obtained from the
steady state solution of Eq.~(\ref{eq:master}) with the rates
specified by Eqs.~(\ref{eq:rate1})-(\ref{eq:rate2}). The
steady-state heat flux calculated, e.g. at the right contact, is
given by the sum
\bea J=\omega_0 \sum_{n=1}^{N-1} n \left( k_R P_n -k_RP_{n-1}
e^{-\beta_R \omega_0} \right),
\label{eq:Curr} \eea
where positive sign indicates current going from left to right.


\subsection{Harmonic molecule}
For the harmonic model ($N \rightarrow \infty $),
putting $\dot{P}_n=0$ in (\ref{eq:master}), 
and searching a solution of the form $P_n \propto y^n$ we get a
quadratic equation for $y$ whose physically acceptable solution is
\beq
 y=\frac{k_L e^{-\beta_L \omega_0} + k_R e^{-\beta_R \omega_0}
}{k_L+k_R}, \label{eq:yy} \eeq
which leads to the normalized state population
\beq P_n=y^n(1-y). \eeq
Using Eq. (\ref{eq:rate2}) we obtain the heat current
(\ref{eq:Curr})
\beq J=\omega_0 \frac{\Gamma_L \Gamma_R}{\Gamma_L+\Gamma_R} \left(
n_L-n_R \right).
\label{eq:J1}
\eeq
In the classical limit, $\omega_0/T_K \ll 1$  ($K=L,R$), it reduces to
\beq J=
\frac{\Gamma_L\Gamma_R}{\Gamma_L+\Gamma_R} \left( T_L-T_R \right).
\label{eq:J1c}
\eeq
This is a special case (with $\mathcal{T}(\omega)= \Gamma_L
\Gamma_R(\Gamma_L+\Gamma_R)^{-1} \delta(\omega-\omega_0) $
consistent with our resonance energy transfer assumption)
\cite{general} of the Landauer expression, Eq.~(\ref{eq:Landauer}).
It is also consistent with the standard expression for the heat current
through a perfect harmonic chain \cite{Lebowitz}.

We emphasize on three important features of this result: (i) The
heat current depends (classically) on the temperature difference
between the two reservoirs, leading to divergent heat
conductivity. Note that there is no need to introduce here the
concept of the molecular temperature $T_M$. (ii) The current is
the same when exchanging $\Gamma_L$ by $\Gamma_R$, i.e.
rectification cannot take place. (iii) The system cannot show the
NDTC behavior, i.e. it is impossible to observe a decrease of the
current with increasing temperature difference. This is true
considering both
 models for the temperature drop- A and B,
(\ref{eq:modelA})-(\ref{eq:modelB}), irrespective of the system
symmetry. We can verify it by studying the $\Delta T$ derivative
of the current (\ref{eq:J1})
\bea \frac{\partial J}{\partial \Delta T}\propto \frac{\partial
n_L}{\partial \Delta T} -  \frac{\partial
n_R}{\partial \Delta T}
=
\frac{\partial n_L}{\partial \Delta T} +  \frac{\partial
n_R}{\partial (-\Delta T)}.
\eea
Since the term
\beq
\frac{\partial n_{L}}{\partial \Delta T}=
\frac { \omega_0 e^{\beta_{L} \omega_0 }}
{T_{L}^2 \left( e^{\beta_{L} \omega_0}-1 \right)^2 }
\frac{\partial T_L}{\partial \Delta T}
\eeq
is always positive (or zero), and similarly the second right hand
side term, NDTC  cannot show up in the fully harmonic
model, and the heat current increases monotonically with the
temperature difference.


\subsection{Anharmonic molecule}

We proceed to the case of a highly anharmonic molecule coupled
-possible asymmetrically- but linearly, to two thermal reservoirs
of different temperatures. We simulate strong anharmonicity by
modeling the molecular mode by a two levels system (TLS). The
Hamiltonian for this model and the resulting rates are
the same as presented throughout Eqs.
(\ref{eq:HT})-(\ref{eq:Curr}), except that we take $n$=0,1 only.
Following Eqs. (\ref{eq:master})-(\ref{eq:kdku}) we obtain the
steady state levels population
\beq P_1=\frac{k_u}{k_u+k_d}; \,\,\,\, P_0=\frac{k_d}{k_u+k_d}.
\label{eq:P12} \eeq
We substitute it into Eq. (\ref{eq:Curr}) with $N$=2 and find the heat current
\cite{Rectif}
\beq J=\omega_0 \frac{\Gamma_L \Gamma_R
(n_L-n_R)}{\Gamma_L(1+2n_L)+\Gamma_R(1+2n_R)}.
\label{eq:J2}
\eeq
Next we calculate the molecular temperature $T_M$ in the weak
coupling-TLS case by substituting the population
(\ref{eq:P12}) into Eq. (\ref{eq:TM}) using Eqs.
(\ref{eq:rate1})-(\ref{eq:rate2}). In the classical limit this
results in
\beq T_M= \frac{\Gamma_L T_L +\Gamma_R T_R}{\Gamma_L+\Gamma_R}
\label{eq:TM1}. \eeq
We can now study the implications of the different models for the
temperature bias, Eqs. (\ref{eq:modelA})-(\ref{eq:modelB}), on the
conductance: In Model A the molecular temperature decreases
monotonically with the temperature difference
\beq T_{M}^{(A)}=T_s-\Delta T \frac{\Gamma_R}{\Gamma_L+\Gamma_R}.
\label{eq:TMA} \eeq
In Model B we find
\beq T_M^{(B)}=T_s+\frac{\Delta T}{2}
\frac{\left(\Gamma_L-\Gamma_R\right)}{\Gamma_L+\Gamma_R},
\label{eq:TMB} \eeq
which implies that for a symmetric ($\Gamma_L=\Gamma_R$) system,
the molecular temperature is constant, whereas in the asymmetric
situation it can either increase or decrease with $\Delta T$, depending on the
sign of $\Gamma_L-\Gamma_R$.

In terms of the molecular temperature (Eq. (\ref{eq:TM1})), going
into the classical limit, the heat current (\ref{eq:J2}) reduces into
the simple form
\beq
 J= (T_L-T_R) \frac{\Gamma_L \Gamma_R}{\Gamma_L+\Gamma_R} \frac{
 \omega_0}{2T_M}. \label{eq:J2c}
\eeq
This relationship differs from the harmonic expression,
(\ref{eq:J1c}), by its implicit dependence on the internal
molecular temperature. As we show next, this opens up the door for
heat rectification and can also lead to the applicability of the
Fourier law of conduction. If we still try to fit this expression
into the Landauer form (\ref{eq:Landauer}), we find that we have
to define an effective {\it temperature dependent} transmission
coefficient $\mathcal{T}(\omega,T_L,T_R)\propto
1/T_M\delta(\omega-\omega_0)$.

\begin{figure}
 {\hbox{\epsfxsize=80mm \epsffile{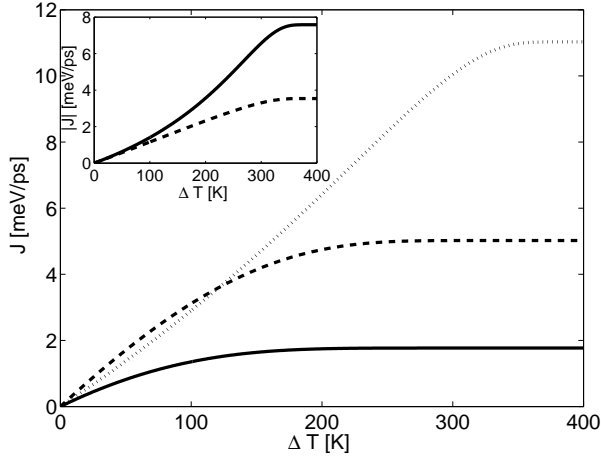}}}
 \caption{
Conduction properties of a TLS system in the weak coupling limit.
 $\omega_0$=150 meV (full), 100 meV (dashed), 25 meV (dotted).
$T_s$=400 K (Model A), $\Gamma_K$=1.2 meV.
Inset: Rectifying behavior of this model,
 $\omega_0$=25 meV, $\chi$=0.75 and $T_L$=400 K, $T_R=T_L-\Delta T$
(full); $T_R$=400 K, $T_L=T_R-\Delta T$ (dashed). }
 \label{FigM2a}
\end{figure}

In Fig. \ref{FigM2a} we display the current (Eq. (\ref{eq:J2}))
for a representative set of parameters. It increases monotonically
with $\Delta T$ and saturates at high temperature gaps. We can
verify this trend analytically as
\bea
\frac{\partial J}{\partial \Delta T}=
 \left[\frac {\partial n_L}{\partial \Delta T}
 (1+2n_R)  - \frac{\partial n_R}{\partial \Delta T}
 (1+2n_L)  \right] \times
\nonumber\\
\frac{\omega_0 \Gamma_L \Gamma_R(\Gamma_L+\Gamma_R)}{\left[ \Gamma_L(1+2n_L) +\Gamma_R(1+2n_R)\right]^2} >0,
\eea
which indicates that NDTC can not take place. However, Eq.~(\ref{eq:J2})
implies that the system can rectify heat current, i.e. the current
can be different (in absolute values) when exchanging the reservoirs
temperatures. Following \cite{Rectif}, defining the asymmetry
parameter $\chi$ such that $\Gamma_L=\Gamma(1-\chi)$ ;
$\Gamma_R=\Gamma(1+\chi)$ with $-1 \leq \chi \leq 1$ we get
\bea \Delta J
&\equiv& J(T_L=T_h; T_R=T_c) +J(T_L=T_c; T_R=T_h)
\nonumber\\
&=&\frac{\omega_0 \Gamma \chi (1-\chi^2)
(n_L-n_R)^2}{(1+n_L+n_R)^2-\chi^2(n_L-n_R)^2}. \label{eq:J2R} \eea
Here $T_c$ ($T_h$) relates to the cold (hot) bath.
Eq.~(\ref{eq:J2R}) implies that for small $\Delta T=T_L-T_R$$,
\Delta J$ grows like $\Delta T^2$, and that the current is larger
(in absolute value) when the cold bath is coupled more strongly to
the molecular system. We exemplify this behavior at the inset of
Fig. \ref{FigM2a}.

We found therefore that a system consisting of an anharmonic
molecular mode coupled linearly (harmonically) and asymmetrically
to two thermal reservoirs of different temperatures can rectify
heat, though it cannot manifest the NDTC effect. NDTC requires
anharmonic interactions with the thermal baths, which may result
in an effective nonlinear temperature dependent molecule-bath
coupling term, see section IV. Therefore, there is no direct
correspondence between these two phenomena.

\subsection{General expression for the heat current}

We can generalize the harmonic (\ref{eq:J1c}) and anharmonic (\ref{eq:J2c})
results and revise the current in the weak coupling limit (${W}$) as
\beq
J_W= \omega_0  \frac{\Gamma_L \Gamma_R}{\Gamma_L+\Gamma_R} \frac{  (T_L-T_R)
}{T_M} f_{S,B}, \label{eq:J2l}
\eeq
where
\bea f_{S,B}=
\begin{cases}
1/2, & \rm { Anharmonic \,\, TLS \,\ case\,\, ( "Spin")} \\
T_M/\omega_0. & \rm { Harmonic\,\, case\,\, ("Boson")} \\
\end{cases}
\eea
For an intermediate anharmonicity we expect this function to
attain an intermediate value, $1/2<f_{S,B}<T_M/\omega_0$. Note
that $f_{S,B}$ can be retrieved by going into the classical limit
of $f_{S,B}=\left[\exp( \omega_0/T_M)\pm1\right]^{-1}$. Here the
"spin" case takes the plus sign, and the "boson" situation
acquires the minus. It can be therefore interpreted as an effective
molecular occupation factor.

We can now clearly trace the influence of the different factors on
the heat conductance. The thermal current is given by  multiplying
three terms: (i) A {\it symmetric} prefactor that includes the
influence of the system-baths coupling, (ii) the factor $\omega_0/
T_M$ which includes internal molecular properties: frequency and
effective temperature, and (iii) the molecular occupation factor
$f_{S,B}$ that varies between $1/2$ for the strictly anharmonic
system and $T_M/\omega_0$ in the harmonic case. As we show next,
the energy current has the same structure when system-bath
interactions are strong.

\section {Strong system-bath coupling}

We turn now to the situation where the  molecular mode  is {\it
strongly} coupled to the thermal reservoirs. As before, we
discuss two limits, the harmonic case, and the anharmonic TLS situation.
In both limits the model Hamiltonian includes the following terms,
as in Eqs. (\ref{eq:HT})-(\ref{eq:H1MB}),
\bea
 {H}&=&
\sum_{n=0}^{N-1}E_n|n\rangle \langle n|
\nonumber\\
&+& \sum_{n=1}^{N-1}\sqrt n V_{n-1,n}|n-1\rangle  \langle n|e^{i
(\Omega_{n}-\Omega_{n-1})}+ c.c.
\nonumber\\
& +& \sum_{j \in L,R} \omega_j
 a_j^{\dagger}a_j,
 \label{eq:H3Tn}
 \eea
where $E_n=n\omega_0$,
 $\Omega_n=\Omega_n^L+ \Omega_n^R$ and $\Omega_n^K=i \sum_{j \in
K} \lambda_{n,j}\left( a_j^{\dagger} -a_j \right)$ ($K=L,R$).
In Appendix A we demonstrate that this model Hamiltonian
equivalently represents a displaced molecular mode coupled
nonlinearly to two thermal reservoirs. The coefficients
$\lambda_{n,j}$  are the effective system-bath interaction
parameters that depend on the level index and the reservoir mode,
see Appendix A. The Hamiltonian (\ref{eq:H3Tn}) is similar to that
defined in Eqs. (\ref{eq:H1M})-(\ref{eq:H1in}), except that the
$L$ and $R$ system-baths couplings appear as multiplicative rather
than additive factors in the interaction term, implying
non-separable transport at the two contacts \cite{Rectif}. The
dynamics is still readily handled. For small $V$ (the
"non-adiabatic limit") the Hamiltonian (\ref{eq:H3Tn}) leads again
to the rate equation (\ref{eq:master}) with
\beq k_d=|V|^2 C(\omega_0);\ \  k_u=|V|^2
C(-\omega_0), \label{eq:kdM3}
 \eeq
where
\beq C(\omega_0)=\int_{-\infty}^{\infty}dt e^{i\omega_0 t}
\tilde{C}(t), \eeq
and
\bea \tilde{C}(t)&=& \left< e^{i[\Omega_{n}(t)-\Omega_{n-1}(t)
]}e^{-i[\Omega_{n}(0)-\Omega_{n-1}(0)]} \right> =\left<
e^{i[\Omega_{n}^L(t) - \Omega_{n-1}^L(t) ]} e^{-i[ \Omega_{n}^L
-\Omega_{n-1}^L] } \right>_L
\nonumber\\
& \times & \left<  e^{i[\Omega_{n}^R(t) - \Omega_{n-1}^R(t) ]}
e^{-i[ \Omega_{n}^R -\Omega_{n-1}^R] } \right>_R.
 \eea
This may be evaluated explicitly to produce
\beq \tilde{C}(t)=\tilde{C}_L(t)\tilde{C}_R(t); \ \
\tilde{C}_K(t)=\exp(-\phi_K(t)),
 \eeq
with
\bea \phi_K(t)=\sum_{j \in K}( \lambda_{n,j}- \lambda_{n-1,j})^2 [
\left( 1+2n_K(\omega_j) \right)
\nonumber\\
 - \left(
1+n_K(\omega_j) \right)e^{-i \omega_j t} -
n_K(\omega_j)e^{i\omega_j t} ].
 \eea
Note that we have omitted the $n$ dependence from the rates above.
This is supported by (i) taking all the inter-levels couplings to
be equal, i.e. $|V_{n-1,n}|=V$, and (ii) assuming that
$(\lambda_{n,j}-\lambda_{n-1,j})^2$ is the same for all $n$ , e.g.
$\lambda_{n,j}\propto n$, see Appendix A.

Explicit expressions may be obtained using the short time
approximation (valid for $\sum_{j\in K}(\lambda_{n,j}-
\lambda_{n-1,j})^2 \gg 1$ and/or at high temperature) whereupon
$\phi(t)$ is expanded in powers of $t$ keeping terms up to order
$t^2$. This leads to
\beq C(\omega_0)=\sqrt{ \frac{2 \pi}{(D_L^2+D_R^2)}} \exp\left[
\frac{-(\omega_0-E_M^L-E_M^R)^2}{2(D_L^2+D_R^2)} \right],
\label{eq:C3}
\eeq
where
\bea E_M^K&=&\sum_{j \in K} (\lambda_{n,j}- \lambda_{n-1,j} )^2
\omega_j, \,\,\,
\nonumber\\
D_K^2&=&\sum_{j \in K} (\lambda_{n,j}- \lambda_{n-1,j} )^2
\omega_j^2\left( 2n_K(\omega_j)+1 \right). \label{eq:DK2} \eea
$E_M^K$ can be considered as the reorganization energy associated
with the structural distortions of reservoirs modes around the
isolated molecular vibration. In the classical limit
($\omega_0/T_K\rightarrow0$), $D_K^2=2T_KE_M^K$.

Following Ref. \cite{Rectif} we calculate the steady state heat
current utilizing
\begin{eqnarray}
J=|V|^2 \sum_{n=1}^{N-1}\int_{-\infty}^{\infty} d\omega \omega [
C_R(\omega)C_L(\omega_0-\omega )P_n
\nonumber\\
-C_R(-\omega)C_L(-\omega_0+\omega)P_{n-1} ]n,
\label{eq:JM3}
\end{eqnarray}
where
\bea
C(\omega_0)&=&\int_{-\infty}^{\infty}d\omega
C_L(\omega_0-\omega)C_R(\omega),
\nonumber\\
C_K(\omega)&=&\frac{1}{\sqrt{2E_M^KT_K}}e^{-(\omega-E_M^K)^2/4T_KE_M^K}.
\eea
Eq. (\ref{eq:JM3}) views the process $|n \rangle
\rightarrow |n-1 \rangle $ in which the molecular mode looses
energy $\omega_0$ as a combination of processes in which the
system gives energy $\omega$ (or gains it if $\omega<0$ ) to the
right bath and energy $\omega_0-\omega $ to the left one, with
probability $nC_L(\omega_0-\omega)C_R(\omega)$. A similar analysis
applies to the process $|n-1\rangle  \rightarrow |n\rangle $.

\subsection{Harmonic molecule}

The levels population of an harmonic molecule ($N\rightarrow
\infty$) are calculated from the steady state solution of Eq.
(\ref{eq:master}), leading to $P_n=y^n(1-y)$, $y=k_u/k_d$, with
the rates conveyed by Eqs. (\ref{eq:kdM3})-(\ref{eq:DK2}). The
heat current (\ref{eq:JM3}) is computed by first making the
summation over $n$
\bea \sum_{n=0}^{\infty} nP_n=
\frac{C(-\omega_0)}{C(\omega_0)-C(-\omega_0)},
\nonumber\\
\sum_{n=0}^{\infty} nP_{n-1} =\frac{C(\omega_0)}{C(\omega_0)-C(-\omega_0)}.
\eea
Next, performing the integrals over frequency yields
\bea J&=&\frac{2\sqrt{\pi} |V|^2  E_M^LE_M^R
(T_L-T_R)}{(E_M^LT_L+E_M^RT_R)^{3/2}}
\nonumber\\
&\times&
 e^{-(\omega_0-(E_M^L+E_M^R))^2/4(E_M^LT_L+E_M^RT_R)} \times f_B,
\label{eq:JC3}
\eea
where
\beq
f_B=\left[e^{\omega_0(E_M^L+E_M^R)/(E_M^LT_L+E_M^RT_R)}-1\right]^{-1}.
\eeq
Before we discuss the heat conduction properties of this model
we examine the anharmonic system.

\subsection {Anharmonic molecule}
The anharmonic model is described by the Hamiltonian
(\ref{eq:H3Tn}) with $n=0,1$. The steady state current is
therefore obtained by reducing Eq.~(\ref{eq:JM3}) to
\begin{eqnarray}
J=|V|^2 \int_{-\infty}^{\infty} d\omega \omega [
C_R(\omega)C_L(\omega_0-\omega )P_1
\nonumber\\
 -C_R(-\omega)C_L(-\omega_0+\omega)P_0 ].
\label{eq:JM4}
\end{eqnarray}
Here $P_0=C(\omega_0)/\left( C(\omega_0) + C(-\omega_0) \right)$
and $P_1=1-P_0$  are established from the steady state solution of
(\ref{eq:master}) with the rates given by (\ref{eq:kdM3}). By
following the same steps  as for the harmonic system, the heat
current (\ref{eq:JM4}) is obtained as \cite{Rectif}
\bea J&=&\frac{2\sqrt{\pi} |V|^2  E_M^LE_M^R
(T_L-T_R)}{(E_M^LT_L+E_M^RT_R)^{3/2}}
\nonumber\\
&\times&
 e^{-(\omega_0-(E_M^L+E_M^R))^2/4(E_M^LT_L+E_M^RT_R)} \times f_S,
\label{eq:JC4}
\eea
with the occupation factor
\beq
f_S=\left[e^{\omega_0(E_M^L+E_M^R)/(E_M^LT_L+E_M^RT_R)}+1\right]^{-1}.
\eeq
%
\begin{figure}
 {\hbox{\epsfxsize=80mm \epsffile{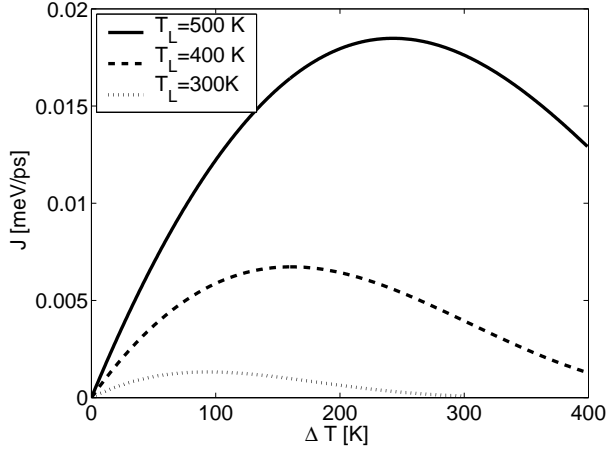}}}
 \caption{
Conduction properties of the TLS system in the strong coupling limit
$T_R=T_L-\Delta T$ (model A),
$E_M^K$=300 meV, ($K=L,R$) $V$=1 meV, $\omega_0$=10 meV.}
 \label{FigM4a}
\end{figure}

\begin{figure}
 {\hbox{\epsfxsize=80mm \epsffile{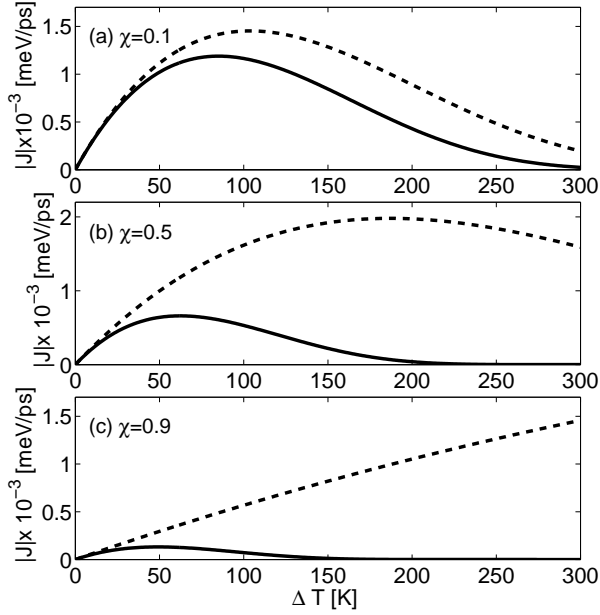}}}
 \caption{Rectification in the strong coupling limit for a TLS system.
$E_M$=300 meV, $V$=1 meV, $\omega_0$=10 meV,
 $T_L$=300 K, $T_R=T_L-\Delta T$ (full); $T_R$=300 K, $T_L=T_R-\Delta T$
  (dashed).}
 \label{FigM4b}
\end{figure}

\subsection{General expression for the heat current}

Next the harmonic (\ref{eq:JC3}) and anharmonic results
(\ref{eq:JC4}) are reduced into a common form. We begin by
evaluating the internal molecular temperature (\ref{eq:TM}). In
the present strong coupling case, for both harmonic and anharmonic
molecular modes, it is given by (\ref{eq:kdM3})
\beq e^{-\omega_0/T_M}\equiv
P_{n+1}/P_n=\frac{C(-\omega_0)}{C(\omega_0)}. \eeq
Using Eq.~(\ref{eq:C3})  we obtain the explicit expression
\beq T_M= \frac{D_L^2+D_R^2}{2(E_M^L+E_M^R)} \stackrel
{\omega_0/T_K \rightarrow 0}{\longrightarrow}
\frac{(E_M^LT_L+E_M^RT_R)}{(E_M^L+E_M^R)}. \eeq
The effective temperature in the strong coupling limit is therefore given by
the algebraic average of the $L$ and $R$ temperatures weighted by
the coupling strengths, here conveyed by the reservoirs reorganization
energies.

In terms of this quantity, we write
a general expression for the current in the strong ($S$) coupling limit as
\bea J_{S}&=& |V|^2 \sqrt {
\frac{4\pi}{T_M(E_M^L+E_M^R)} }
e^{-(\omega_0-E_M^L-E_M^R)^2/4T_M(E_M^L+E_M^R)}
\nonumber\\
&\times& \frac{E_M^LE_M^R}{E_M^L+E_M^R} \frac{T_L-T_R}{T_M} \times
f_{S,B}, \label{eq:JCg} \eea
where
\bea f_{S,B}&\equiv&
\left[e^{\omega_0(E_M^L+E_M^R)/(E_M^LT_L+E_M^RT_R)}\pm1\right]^{-1}
\nonumber\\
&=& \left(e^{\omega_0/T_M} \pm 1\right)^{-1}. \eea
The plus sign relates to the anharmonic "spin" case,  the minus
stands for the harmonic "boson" situation.

We analyze next the conduction properties of this model. Diode
like behavior is expected when $E_M^L\neq E_M^R$, since then the
resulting molecular temperature $T_M$ is not the same when
exchanging $T_L$ with $T_R$. Note that in the present strong
(nonlinear) coupling limit the molecule does not need to be strictly
anharmonic for demonstrating this behavior, in contrast to the weak
coupling situation.

NDTC can also take place in the system,
depending on the system asymmetry and the specific model for the applied
temperature gradient.
When the temperature bias is applied
symmetrically at the $L$ and $R$ sides (model B, Eq.
(\ref{eq:modelB})), NDTC occurs for an {\it asymmetric}
$E_M^L\neq E_M^R$ system. In model A the molecular temperature
depends on $\Delta T$ even for a symmetric junction, providing NDTC.

Figure \ref{FigM4a} depicts an example of NDTC behavior in the
system. The left reservoir is held at a constant temperature,
while the temperature of the $R$ reservoir is decreasing. We find
that up to $\Delta T=T_L-T_R \sim 100 $ K the current increases
with the temperature bias, while above it, i.e. for lower $T_R$,
the current goes down, and even diminishes (dotted line).

We can also investigate the effect of asymmetrical contacts.
We define the asymmetry parameter $\chi$ such as
$E_M^L=E_M(1-\chi)$, $E_M^R=E_M(1+\chi)$,  $0<\chi<1$. Figure
\ref{FigM4b} presents the heat current  when
$\chi \neq 0$. (a) For small $\chi$ the current is almost the same
for both forward and reversed operation modes. (b) At
intermediate $\chi$ values we find that for $T_L=100$ K, $T_R$=300 K
there is a maximal heat flow (dashed), while for the reversed
operation ($T_R=100$ K, $T_L$=300 K)  heat current is blocked
(full). (c) For a highly asymmetric system heat flows predominantly
in one direction.


We can further formulate a general expression for the current that
holds in {\it both} strong and weak interaction regimes and for
either harmonic or anharmonic systems. For convenience, we copy
here the weak ($W$) {\it linear} coupling result (\ref{eq:J2l})
\beq
J_{W}= \omega_0  \frac{\Gamma_L \Gamma_R}{\Gamma_L+\Gamma_R} \frac{  (T_L-T_R)
}{T_M} f_{S,B}.
\eeq
Comparing it to Eq. (\ref{eq:JCg}) guides us to the compact
expression
\beq J=C \frac{f_{S,B}}{T_M}  \Delta T .
\label{eq:JG}
\eeq
Here $C$ includes the contact contribution, which is different in
the weak and strong coupling regimes. It may depend on the
molecule-baths microscopic couplings, molecular vibrational
frequency and the reservoirs temperatures. It is not influenced by
the degree of molecular harmonicity which affects only the
"Spin-Boson" factor $f_{S,B}$. The  temperature  $T_M$ provides
the effective temperature of the molecular system that is
irrelevant in the fully harmonic case. We can therefore clearly
distinguish in this expression between the role of the system
harmonicity and the effect of molecule-baths interactions.
%


\begin{figure}
 {\hbox{\epsfxsize=90mm \epsffile{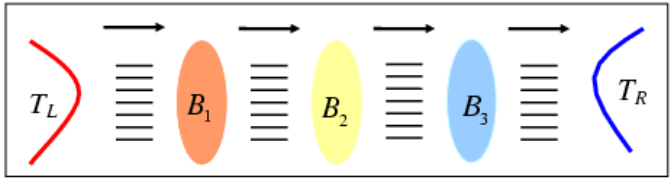}}}
\vspace{0mm} \caption{(Color online) A schematic representation of
the $l$=4 molecular modes system with three intermediate thermal
baths.} \label{scheme}
\end{figure}

\section{Fourier law of conduction}

The validity of Fourier's law of heat conduction
(\ref{eq:Fourier}) in 1D lattices is an open issue \cite{Prosen}.
This law is a macroscopic consequence of ordinary diffusion at the
microscopic level. Here we extend our single mode model and
present a realization of a molecular chain that can lead to normal
Fourier conduction, provided that the molecule is highly
anharmonic, independently of the molecule-baths coupling
strengths. We emphasize that we bypass the difficult task of
showing how normal diffusion emerges in the system \cite{Schr},
and simply assume non-correlated hopping motion between molecular
units. Our sole mission here is to construct from the single mode
result (\ref{eq:JG}) a model that supplies normal conduction.

Figure \ref{scheme} depicts the system: We envision an array of
$I+2$ local heat baths where heat transfers along the chain of
bath, single mode, bath, single mode...: $L\rightarrow B_1
\rightarrow B_2 \rightarrow ... \rightarrow B_I \rightarrow R$.
The intermediate thermal baths $B_i$ might  be realized by large
molecular groups where full thermalization to a local temperature
occurs. This implies that anharmonic interactions govern the
dynamics within the intermediate baths. We also assume that the
temperature difference $T_L-T_R$ falls linearly on the system,
i.e. the temperature at the $i$ bath is
\bea T_i=T_s+\delta T[i-(I+1)/2],\,\,\ \Delta T=\delta T (I+1),
\eea
where $i=(0,1,2...I,I+1)$. The $L$ and $R$ bath are indiced by 0
and $I+1$ respectively, and $\delta T$ is the  temperature
difference between neighboring reservoirs. The resulting effective
temperature of a molecular mode located in between each two baths
$i-1$ and $i$ is
\beq T_M^{(i)}=T_s+\delta T[2i-1-(I+1)]/2,
\eeq
assuming equal coupling along the chain ($\Gamma_i=\Gamma$ or
$E_M^i=E_M$).

The  conductance $\kappa_i$ of the unit $B_{i-1}\leftrightarrow
B_{i}$ is defined through the relation
$J=\kappa_i(T_{i}-T_{i-1})$. In the diffusional hopping regime,
the total conductance $\kappa_T$ is established by inversing the
sum of all units resistances, $\kappa_T=\left(\sum_i
1/\kappa_i\right)^{-1}$. In the linear coupling- TLS model
(\ref{eq:J2l}) it becomes
\beq \kappa_T=\frac{\omega_0\Gamma
}{4}\left(\sum_{i=1}^{I+1}T_M^{(i)}\right)^{-1}= \frac{\Gamma
\omega_0}{4T_sl},
\eeq
where $l=I+1$ is the actual length of the system, given by the
number of internal molecular modes. This equation implies the
validity of the Fourier law of heat conduction, $J\propto
(T_L-T_R)/l$. In addition, the prefactor depends on the inverse
temperature $T_s^{-1}$ as expected in the high temperature limit
\cite{Kittel} where interactions among phonons are dominant.
The same result holds when the temperature falls mainly on the contacts
whereas the temperature distribution along the central baths is almost constant.
Strong molecule-bath interactions  contribute a
temperature dependent prefactor, but do not modify the $l^{-1}$ form assuming
$\Delta T/T_s \ll 1$.
A possible realization of this system is a fullerene polymer \cite{fullerene}.

\section{Conclusions}

\begin{figure}
{\hbox{\epsfxsize=70mm \epsffile{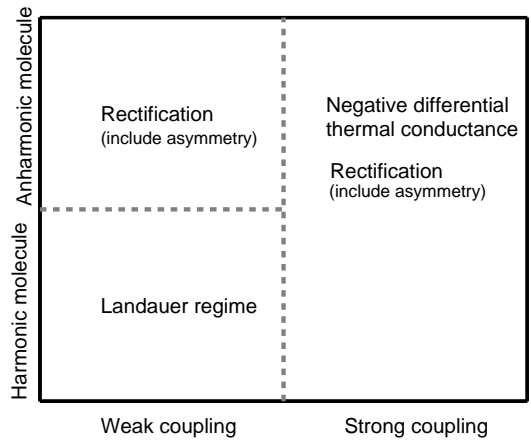}}}
\caption{An overview of the parameter ranges providing negative differential
thermal conductance (NDTC) and diode like behavior. The $x$ axis,
(weak and strong coupling) relates to the system-bath interaction model.
In order to obtain rectification the system should be asymmetric with respect
to the $L$ and $R$ ends.}
\label{Figo}
\end{figure}

Using a simple theoretical model we have investigated the effect
of anharmonic interactions on heat flow through molecular
junctions. Our general expressions for the heat current
(\ref{eq:J2l}) and (\ref{eq:JCg})  clearly manifest the interplay
between the system anharmonicity, system-bath coupling and
junction asymmetry. We have also extended our model into a chain
of $l$ local molecular sites, and indicated that anharmonicity is
a crucial requirement for achieving normal conductance.
We have found that nonlinear interactions can lead to novel
phenomena: negative differential thermal conductance and heat
rectification. NDTC takes place when the molecular mode is
strongly coupled to the thermal environment. In contrast, diode
like behavior originates from the combination of
 substantial molecular anharmonicities with a structural asymmetry.
Figure \ref{Figo} presents an overview of
the different regimes studied, and the nonlinear effects observed in each case.

We would also like to draw an analogy between the nonlinear
behavior discussed in this paper and some nonlinear effects
discovered in molecular level {\it electron} carrying systems: The
negative differential resistance observed in molecular films of
C$_{60}$ could be explained due to a voltage dependent tunneling
barrier \cite{Crommie}.  Rectification of electron current was
theoretically exhibited in one-dimensional asymmetric electronic
conductors with screened electron-electron interactions
\cite{Brend}.

Control of heat flow through molecules by employing nonlinear
interactions might be useful for different applications. In
molecular electronic local heating of nanoscale devices might
cause structural instabilities undermining the junction integrity
\cite{Heat}. Engineering good thermal contacts and cooling of the
the junction \cite{SegalPump} are necessary for a stable and
reliable operation mode. Control of vibrational energy transfer in
molecules affects chemical processes, e.g. reaction pathways, bond
breaking processes, and folding dynamics \cite{Leitner2}. Finally,
we propose building technological devices based on heat flow, in
analogy with electron current devices \cite{CasatiR}.

\begin{acknowledgments}
The author would like to thank Professor A. Nitzan for his comments.
This project was supported by the Feinberg graduate school of
the Weizmann Institute.
\end{acknowledgments}


\renewcommand{\theequation}{A\arabic{equation}}
\setcounter{equation}{0}  
\section*{APPENDIX A: Microscopic model for the strong coupling Hamiltonian }

The strong coupling Hamiltonian Eq.~(\ref{eq:H3Tn}) can be derived
from the following microscopic model
\begin{eqnarray}
 H&=& \sum_{n=0}^{N-1}n\omega_0|n\rangle \langle n|+
\sum_{j \in L,R}\frac{\omega_j^2}{2}\left( x_j- \sum_{n=0}^{N-1}\frac{
n\alpha_{n,j}}{\omega_j^2}|n\rangle \langle n| \right)^2
\nonumber\\
&+& \sum_{n=1}^{N-1}\sqrt{n}V_{n,n-1}|n\rangle \langle n-1|
 + \sqrt{n}V_{n-1,n}|n-1\rangle \langle n|
\nonumber\\
 &+&\sum_{j \in L,R} \frac{p_j^2}{2},
 \label{eq:AH3T}
 \end{eqnarray}
which describes a forced oscillator of frequency $\omega_0$
strongly interacting with the $L$ and $R$ thermal baths. The
nonlinear contributions are concealed in the second element of
(\ref{eq:AH3T}) providing high order terms such as $\propto x_j
x^2$, with $x$ as the molecular coordinate. Here $x_j$ and $p_j$
are the displacement and momentum of the reservoirs harmonic modes
with frequency $\omega_j$, $\alpha_{n,j}$ is the system-bath
coupling parameter and $V_{n,n-1}$ is the effective inter-level
matrix element. We can expand the quadratic term in
(\ref{eq:AH3T}) and obtain
\begin{eqnarray}
 H&=& \sum_{n=0}^{N-1}n\left( \omega_0-  \sum_{j\in L,R}x_j\alpha_{n,j}\right)
|n\rangle\langle n|
\nonumber\\
&+& \sum_{n=1}^{N-1}\sqrt{n}V_{n,n-1}|n\rangle \langle n-1|
 + \sqrt{n}V_{n-1,n}|n-1\rangle \langle n|
\nonumber\\
 &+&\sum_{j \in L,R} \omega_j a_j^{\dagger}a_j
+\sum_{n=0}^{N-1}\sum_{j\in L,R}\frac{n^2\alpha_{n,j}^2}{2\omega_j^2}|n\rangle
\langle n|.
 \end{eqnarray}
Here
$x_j=(a_j^{\dagger}+a_j)/\sqrt{2 \omega_j}$
and $p_j=i\sqrt{\omega_j/2}(a_j^{\dagger}-a_j)$.
Use of the small polaron transformation \cite{Mahan},
$\tilde{H}= UHU^{-1}$, leads to
\bea
 \tilde{H}&=&
\sum_{n=0}^{N-1}n\omega_0|n\rangle \langle n| +\sum_{n=0}^{N-1}\sum_{j\in
L,R}\frac{n^2\alpha_{n,j}^2}{2\omega_j^2}|n\rangle \langle
n|+H_{shift},
\nonumber\\
&+& \sum_{n=1}^{N-1}\sqrt n V_{n-1,n}|n-1\rangle  \langle n|e^{i
(\Omega_{n}-\Omega_{n-1})}+ c.c.
\nonumber\\
& +& \sum_{j \in L,R} \omega_j
a_j^{\dagger}a_j,
\label{eq:AH3Tn}
\eea
where
\beq U=\Pi_{n=0}^{N-1} U_n, \,\,\
U_n=\exp(-i\Omega_n|n\rangle \langle n|),
\eeq
and where
\bea
\Omega_n&=&\Omega_n^L+ \Omega_n^R, \,\,\ \Omega_n^K=i \sum_{j
\in K} \lambda_{n,j}\left( a_j^{\dagger} -a_j \right),\,\ (K=L,R),
\nonumber\\
\lambda_{n,j}&=&(2\omega_j^3)^{-1/2}n\alpha_{n,j}.
\eea
The term
\beq
H_{shift}=-\frac{1}{2}\sum_{n=0}^{N-1}\sum_{j}\frac{ n^2 \alpha_{n,j}^2}{\omega_j^{2}}
|n\rangle \langle n|
\eeq
exactly cancels the $\propto n^2$ term in Eq. (\ref{eq:AH3Tn}),
and we finally recover the strong coupling Hamiltonian (\ref{eq:H3Tn}).



\end{document}